# Zero Field Cooled Exchange Bias Effect in Nano-Crystalline Mg-Ferrite Thin Film

Himadri Roy Dakua

Department of physics, Indian Institute of Technology Bombay, Mumbai, India 400076

Email: hroy89d@gmail.com

**Abstract**

I report, Zero Field Cooled (ZFC) Exchange Bias (EB) effect in a single phase nanocrystalline Mg-ferrite thin film, deposited on an amorphous quartz substrate using pulsed laser ablation technique. The film showed a high ZFC EB shift ($H_E \sim$ 190 Oe) at 5 K. The ZFC EB shift decreased with increasing temperature and disappeared at higher temperatures (T > 70 K). This Mg-ferrite thin film also showed Conventional Exchange Bias (CEB) effect, but unlike many CEB systems, the film showed decrease in the coercivity ($H_C$) under the Field Cooled (FC) measurements. The film also showed training effect in ZFC measurements which followed the frozen spin relaxation behaviour. The observed exchange bias could be attributed to the pinning effect of the surface spins of frozen glassy states at the interface of large ferrimagnetic grains.

## I.    Introduction

Meiklejohn and Bean discovered the Exchange Bias (EB) effect in the Ferromagnetic (FM) Co and Antiferromagnetic (AFM) CoO core-shell heterostructures.[1] The exchange bias effect was characterized by a horizontally (along the field axis) off-centred M-H loop of the Field Cooled (FC) core-shell heterostructure. Since then, a lot of research have been carried out on different exchange bias systems due to their applicability in many magnetic devices such as data storage devices, spin valve devices and voltage control magnetic devices etc.[2-6]

Apart from these Conventional Exchange Bias (CEB) systems, Wang *et al.* reported an unusual exchange bias effect in the $Ni_{50}Mn_{50-x}In_x$ (x = 11-15) bulk Heusler alloys in 2011.[7] Here, unlike the CEB systems, the sample showed a large shift in the M-H loops even in the Zero Field Cooled (ZFC) measurements. Subsequently, Nayak *et al.* also reported ZFC Exchange Bias (EB) effect in bulk Heusler alloy $Mn_2PtGa$.[8] Recently, some other groups also reported ZFC EB effect in few more bulk materials.[9-11] All these studies had broadly pointed out that a unidirectional anisotropy is introduced to the system during the initial magnetization process. While, the microscopic origin of the ZFC EB effect is not yet fully understood.

This paper focuses on the exchange bias effect in Mg-ferrite nanocrystalline thin film. The cubic spinel ferrites such as Mg, Ni, Mn –ferrites are well known magnetic materials for the high frequency applications.[12,13] The ferrimagnetic ordering in these ferrite systems is mainly due to the anti-parallel alignment of cation spins at the tetrahedral (A) and the octahedral (B) sites. The chemical formula of these cubic spinel ferrites is expressed as $(M_{1-x}Fe_x)_A(M_xFe_{2-x})_BO_4$ based on their cation occupancy.[14] In Mg-ferrite bulk sample, a (x = ~ 0.89) faction of $Fe^{3+}$ ions occupy the A sites while other (2-x) in the B sites and this leads to the ferrimagnetic ordering in it.[14,15] However, it is to be noted that these single phase bulk spinel ferrites ($MFe_2O_4$, M = Mg, Mn, Co, Ni) do not show exchange bias effect. Though, there are few reports on conventional exchange bias effect in thin films of some ferrites. Like, Venzke *et al.*[16] observed the CEB effect in as deposited Ni-ferrite thin films. Alaan *et al.* also reported exchange bias effect in MnZn - ferrite thin films.[17,18] While in case of Mg-ferrite thin films, some inconsistent and self-contradicting data on exchange bias effect were also reported earlier.[19,20] Therefore, the details and true behaviour of the exchange bias effect in Mg-ferrite thin films are still unknown. Here, I have presented the detail study of the exchange effect in Mg-ferrite thin film. The data presented in this paper, shows some distinguishably deferent features compared to the CEB effect. These features are compared with the other exchange bias systems and discussed in this paper.

## II. Experimental details
### a. Details of the thin film growth conditions

Nanocrystalline Mg-ferrite thin film was deposited using pulsed laser ablation technique. A single phase high density Pulsed Laser Deposition (PLD) target was prepared through solid state reaction route. The film was deposited using a Nd:YAG pulsed laser with energy density 2 Joule/cm². The pulsed laser repetition rate was kept at 10 shots/sec and the film was deposited on quartz substrate using 18000 pulsed laser shots. The clean amorphous quartz substrate was kept 4.5 cm away from the PLD target and was heated to 500 °C while taking the deposition. The deposited film was ex-situ annealed at 250 °C for 2 hrs in air and cooled down to room temperature (RT) through atmospheric cooling in closed furnace. All the measurements were performed using this annealed film.

### b. Magnetization loops (M-H) measurement details

The field dependence of magnetization (M-H) of the film was measured using two protocols, ZFC and FC. For the ZFC measurements, the film was cooled down in zero magnetic field from RT. The ZFC M-H loops were collected by sweeping the magnetic field in two different ways. In the first way (p-type), the field was swept from 0 Oe → +50 kOe → -50 kOe → +50 kOe . In the second way (n-type), the field was swept from 0 Oe → -50 kOe → +50 kOe → -50 kOe. In these measurements the initial 0 Oe → ±50 kOe, magnetization curve is termed as virgin M-H curve.

The FC M-H loops were collected by cooling the film in an applied field $H_{FC}$, from RT and the M-H loops were measured by sweeping the field as $H_{FC}$ → -50 kOe → +50 kOe. Prior to all the measurements the film was subjected to a damped oscillating magnetic field (centred at 0 Oe) which gradually becomes zero at RT. This process ensured the zero magnetization state of the film at RT. All the measurements were performed by applying the magnetic field along the film's plane.

## III. Results
### a. Structural and elemental properties of the film

Fig. 1 shows the GIXRD data of the film measured at room temperature with an incident angle 0.5°.

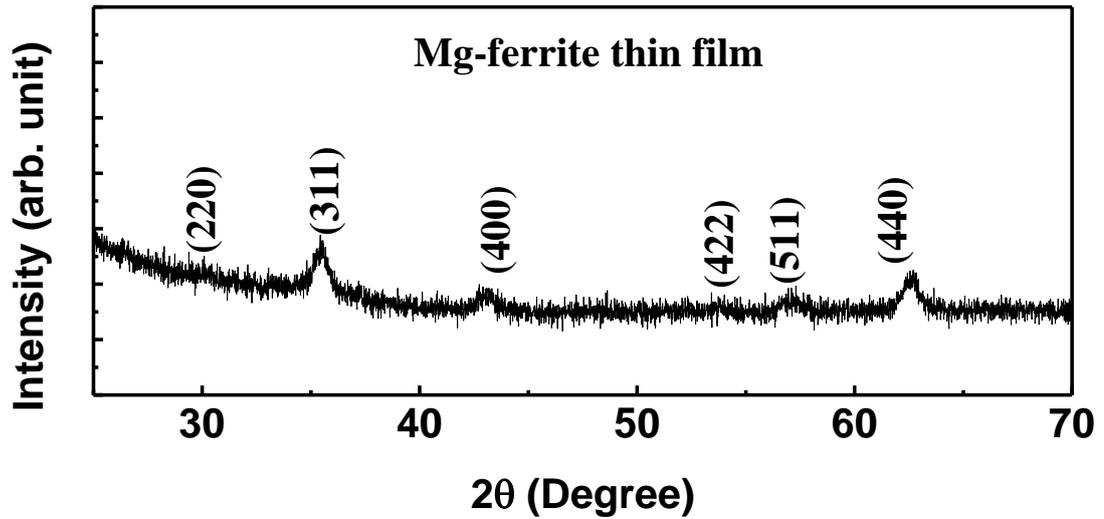

Fig. 1. GIXRD of the magnesium ferrite thin film, deposited at $T_S$ = 500 °C and annealed at 250 °C for 2 hours

The GIXRD of the film shows diffraction peaks correspond to the cubic spinel structure of Mg-ferrite of space group Fd3m. The observed peaks are identified and indexed in the Fig. 1.

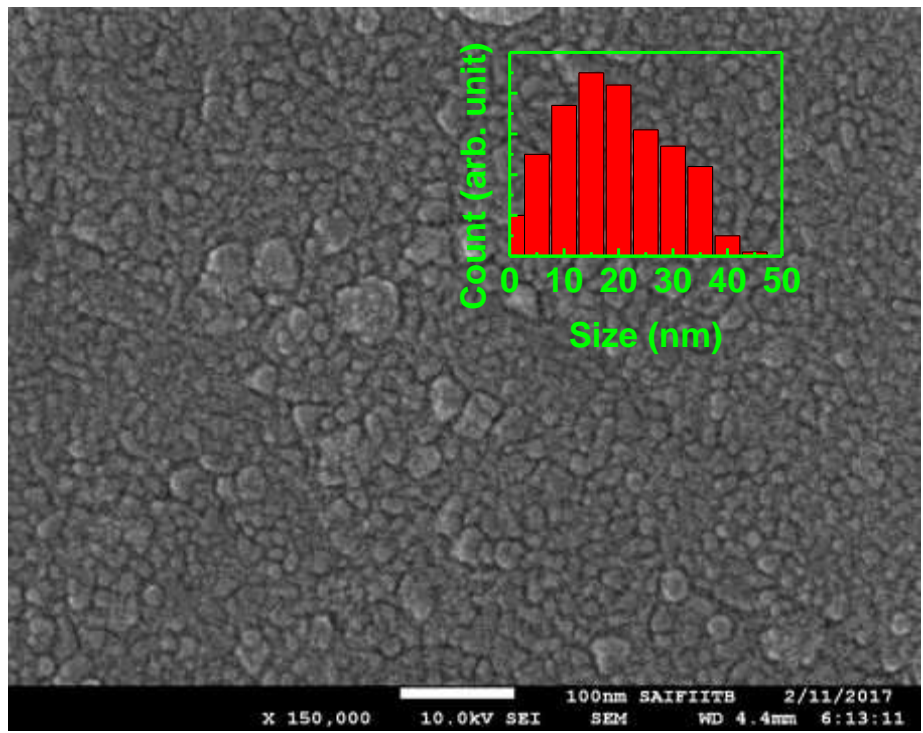

Fig. 2. Planner FEG-SEM image of the Mg-ferrite thin film

Fig. 2 shows the FEG-SEM image of the film surface. The FEG-SEM shows nano size grains. The size of these grains were measured and the histogram of the size distribution is shown in the inset of the Fig. 2. The grain size varies from few nanometre to ~ 40 nm with a distribution peak at ~ 17 nm. The thickness of the film was around 135 nm.

The elemental analysis of the film was performed using X-ray Photoelectron Spectroscopy (XPS) data. Fig. 3 (a) shows the XPS survey profile of the film. The observed peaks can be identified due to the Magnesium (Mg), Iron (Fe), Oxygen (O) and surface absorbed Carbon (C) XPS and Auger peaks. The observed peaks are indexed in the figure. The survey spectra do not show any additional peak correspond to alien elements, which confirms the elemental purity of the deposited film. Fig. 3 (b), (c) and (d) show the high resolution core level XPS spectra of Mg 1s, Fe 2p and O 1s respectively. The Mg 1s core level spectra shows peak at 1303.08 eV which is similar to that observed in $MgFe_2O_4$ system by *Mittal et al.*[21]

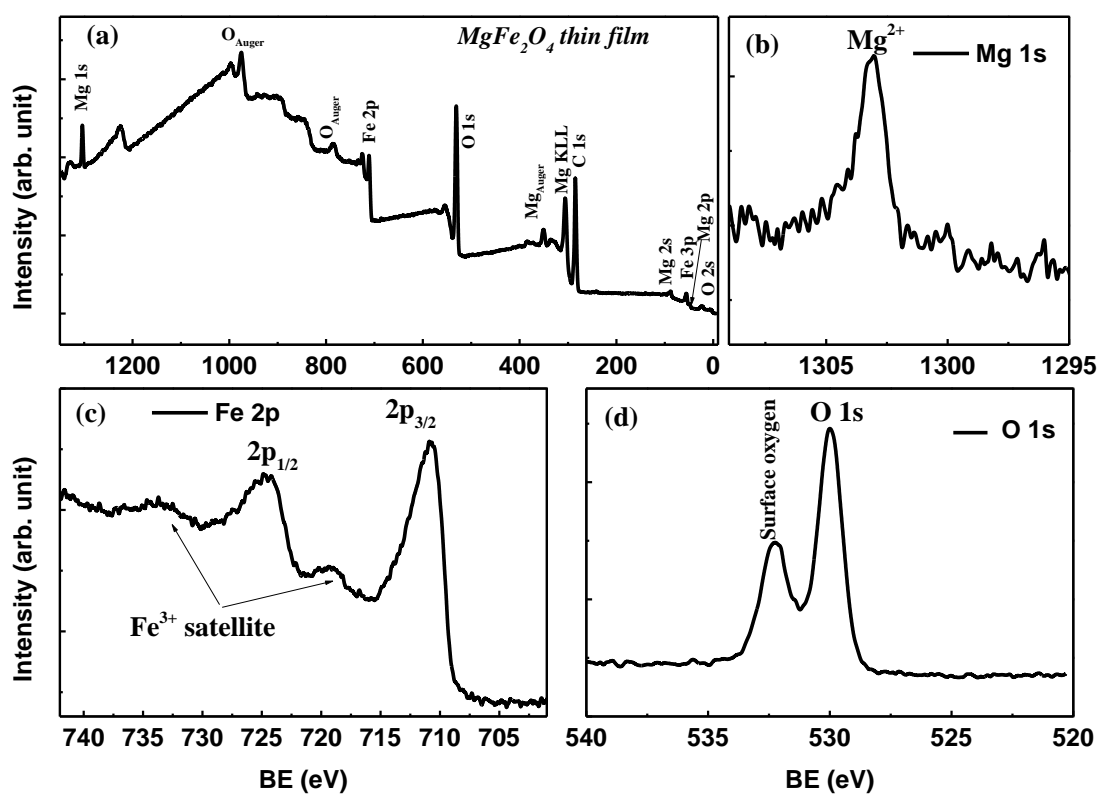

Fig. 3. (a) XPS survey spectroscopy of the film, (c) Mg 1s, (c) Fe 2p and (d) O 1s high resolution core level spectra.

The Fe 2p core level spectra shows two satellite peaks at 719.5 eV and ~733.6 eV. These satellite peaks confirm the presence of $Fe^{3+}$ ionic state in the system. A similar Fe 2p spectra was also obtained for $Fe^{3+}$ ions by other research groups.[21-23] The oxygen 1s core level spectra shows two peaks (at 530.2 eV and 532.3 eV) correspond to surface absorbed oxygen (532.3 eV)[24] and 1s core level spectra (530.2 eV) of oxygens of the $MgFe_2O_4$. The GIXRD, FEG-SEM and the XPS results suggest that the film is single phase nano-crystalline and impurity free $MgFe_2O_4$.

### b. Zero field Cooled (ZFC) and Field Cooled Exchange Bias effect

The Mg-ferrite thin film showed Exchange Bias (EB) effect even in Zero Field Cooled (ZFC) measurements. Here I present the detail features of this ZFC EB effect. Fig. 4 (a) shows the p-type ZFC M-H loop of the film, measured at 10 K. The open circle data represent the virgin M-H curve. Fig. 4 (b) shows the zoomed view of the Fig. 4 (a). This figure clearly shows that the ZFC M-H loop is shifted towards the negative field axis. The observed shift is termed as ZFC Exchange Bias (EB) shift (or field) and measured as $H_E = |(H_{C1} + H_{C2})/2|$, where $H_{C1}$ and $H_{C2}$ are the two intercepts of the magnetization curve with the field axis as shown in Fig. 4 (b). The average coercivity of the M-H loop is measured as $H_C = |H_{C1} - H_{C2}|/2$. It was also observed that in this p-type ZFC M-H loop, the value of the remanence magnetization $|M_{r1}|$ (= 73.4 emu/cc) on positive y-axis is smaller than that of the negative y-axis, $|M_{r2}|$ (= 81.6 emu/cc). The average remanence magnetizations is expressed as $M_r = \frac{|M_{r2}| + |M_{r1}|}{2}$.

Another distinguishable feature is observed in the virgin magnetization curve of the ZFC M-H loop. The Fig. 4 (a) clearly shows that a portion of virgin magnetization curve (open circle) is outside the M-H loop. This curve exited the loop at a magnetic field H' and merged with the M-H data at a magnetic field H". A similar behaviour of the virgin magnetization curve was also reported in some bulk ZFC EB systems.[7, 8, 25] This behaviour was speculated due to a field induced ordering in the system.[7, 9]

Here one need to note that the shifted asymmetric M-H loops were also observed not only due to the exchange bias effect but also due to minor loop and experimental artefacts.[26] A minor M-H loop generally shows vertical shift, open loop and non-saturation.[26] In case of this film, it was observed that the magnitude of the high field magnetizations ($|M_{+50\,kOe}|$ and $|M_{-50\,kOe}|$) of the M-H loop are equal and also the hysteresis in the M-H loop disappeared in the high field region (|H| > 30 kOe). These confirmed that this M-H loop is not a minor loop of the film.

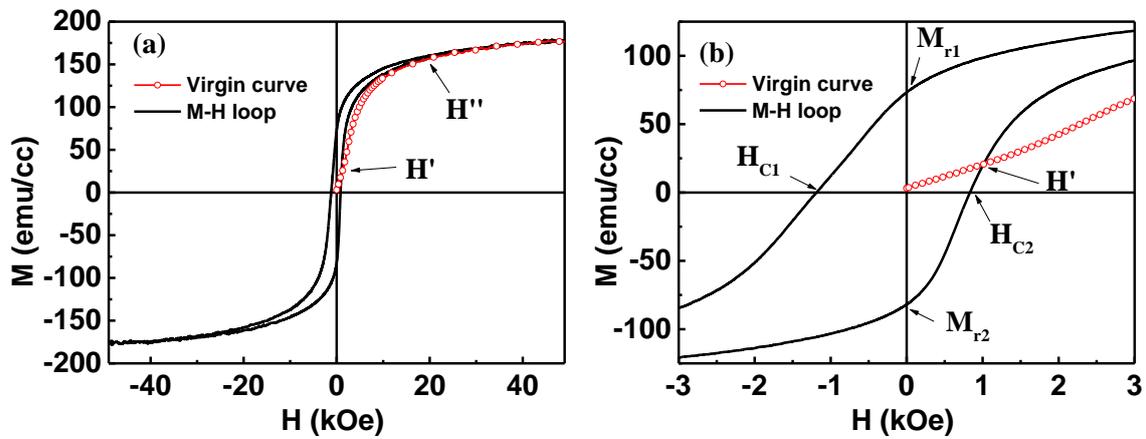

Fig. 4. (a) 10 K ZFC p-type M-H loop. The virgin M-H curve is shown in red open circle. (b) Shows the zoomed view of the figure (a).

The possibility of measurement artefacts in the ZFC EB effect was checked by measuring the ZFC p-type and n-type M-H loops. Previously, similar procedure was also followed by different groups to check the measurement artefacts in EB effect. [7, 25]

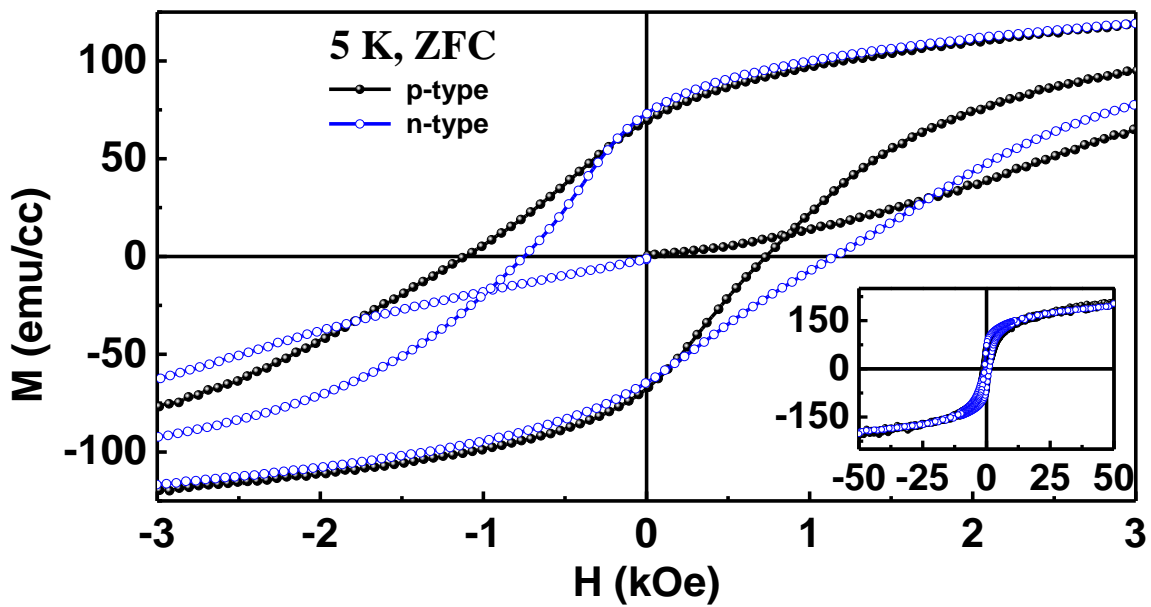

Fig. 5. K ZFC p-type and n-type M-H loops, plotted within the field range ±3 kOe. Inset shows these ZFC p-type and n-type loops within the field range ±50 kOe.

Fig. 5 shows the p-type and n-type ZFC M-H loops along with the corresponding virgin magnetization curves of the film, measured at 5 K. The high field magnetizations (both |M$_{+50 kOe}$| and |M$_{-50kOe}$|) of the film remain same irrespective of these two different measurements. But the shift along the magnetic field axis has changed. The p-type M-H loop shifted along the negative field axis (H$_E$ = 190 Oe), whereas, the n-type M-H loop shifted along the positive field axis (H$_E$ = 198 Oe). The exchange bias shift observed in the both p-type and n-type measurements were found to be equivalent and opposite. Moreover, the value of │H'│ for both the p-type (H' = 912 Oe) and n-type (H' = -947 Oe) loops were found to be similar. These results indicate that the observed EB effect is due to Zero Field Cooled Mg-ferrite film and not due to any experimental artefacts.[7, 25]

I have also studied the Conventional Exchange Bias (CEB) effect in this Mg-ferrite thin film. Fig. 6 shows the ZFC p-type and FC M-H loops measured at 10 K. The FC measurements were performed after cooling down the film in presence of a field, H$_{FC}$, (here H$_{FC}$ = 50 kOe). Similar to the ZFC p-type M-H loop, the FC M-H loop also showed exchange bias shift along the negative field axis.

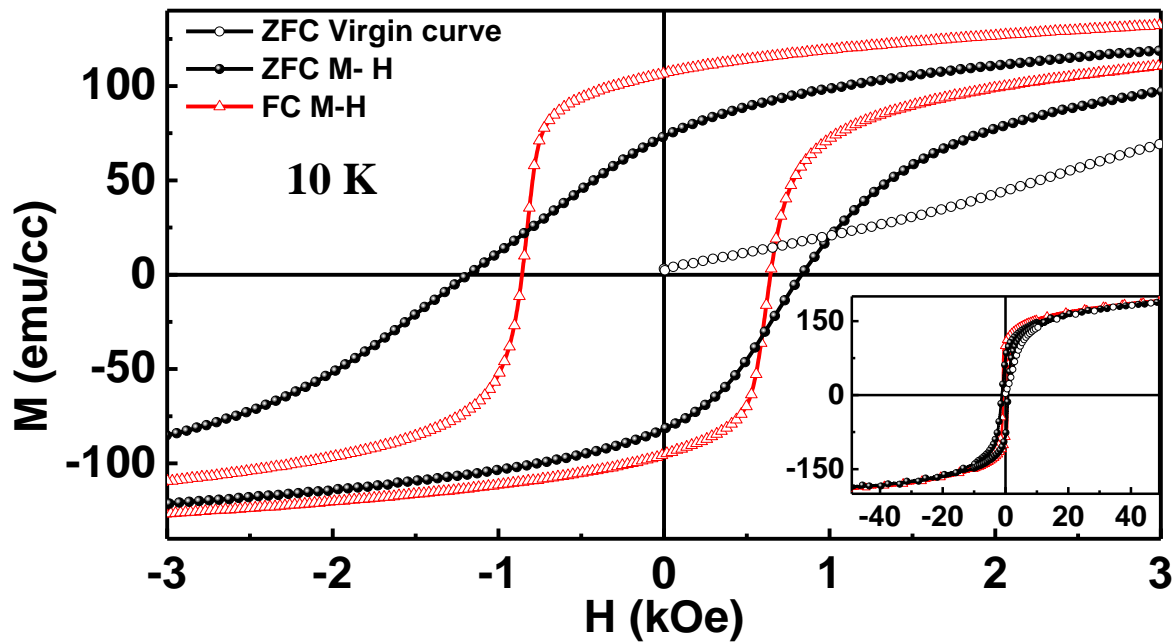

Fig. 6. ZFC p-type and FC M-H loops, measured at 10 K, the loops are enlarged within ±3 kOe. Inset: full range (± 50 kOe) ZFC and FC M-H loops measured 10 K. The open circles are virgin M-H curve of the ZFC measurement.

The FC EB shift was found to be $H_E$ = 110 Oe for $H_{FC}$ = 50 kOe, which is lower than the ZFC EB field, $H_E$ = 177 Oe. The coercivity ($H_C$ = 750 Oe) of this FC M-H loop is also smaller than the ZFC $H_C$ (= 1000 Oe). One needs to note that this behaviour is not common in CEB systems. The CEB systems generally show an enhancement in the coercivity of the field cooled M-H loops.[27, 28]

Fig. 6 also shows that the values of the $|M_{r1}|$ (= 108 emu/cc) and $|M_{r2}|$ (= 96 emu/cc) of the 50 kOe FC M-H loop are higher than that of the $|M_{r1}|$ (= 73 emu/cc) and $|M_{r2}|$ (= 82 emu/cc) of ZFC p-type M-H loop, respectively. Whereas, the high field ($\pm$ 50 kOe) magnetization of both the FC and ZFC M-H loops were same. The increase in remanence magnetization of FC M-H loops of CEB systems is also reported but it often associated with an equivalent change in high field magnetization too.[29-32]

The temperature and cooling field ($H_{FC}$) dependence of the exchange bias shift of the Mg-ferrite thin film was also studied. Fig. 7 (a) shows the temperature dependence of the exchange bias field, $H_E$, for $H_{FC}$ = 0 Oe and 50 kOe. Fig. 7 (b) shows the temperature dependence of coercivity ($H_C$) for $H_{FC}$ = 0 and 50 kOe. The film showed higher value of $H_E$ for $H_{FC}$ = 0 Oe (ZFC) than that of the $H_{FC}$ = 50 kOe. The ZFC and the 50 kOe FC exchange bias shift decreased with increasing temperature.

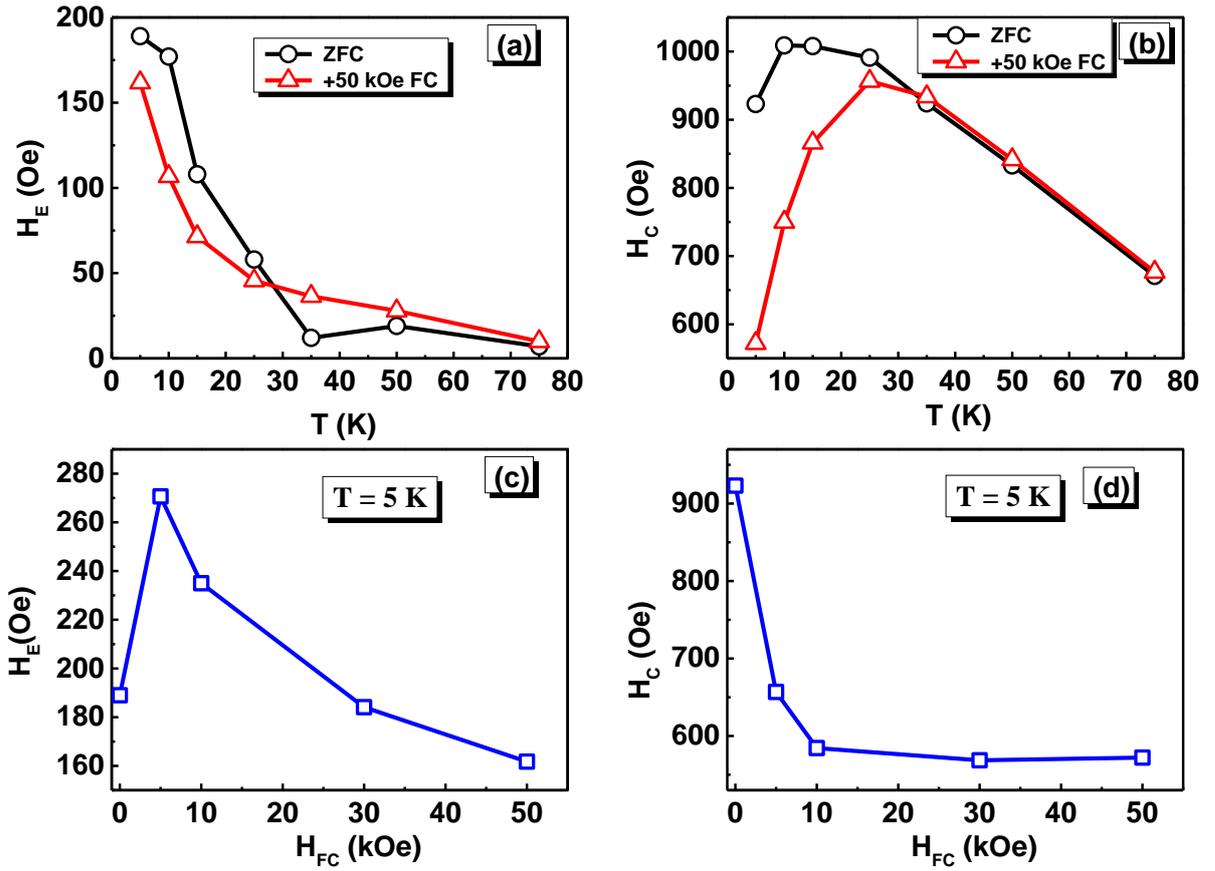

Fig. 7. (a) Temperature dependence change in ZFC (open circle), FC (open Tringle) exchange bias field $H_E$ (b) Temperature dependent coercivity of the ZFC and FC M-H loops. The cooling field ($H_{FC}$) dependent (c) exchange bias shift ($H_E$) and (d) coercivity ($H_C$) at 5 K. Lines are guide to eyes.

The coercivity $H_C$, of the ZFC M-H loops decreased almost monotonically as the temperature increased (except the 5 K data). While, the coercivity ($H_C$) of the 50 kOe FC M-H loops showed a smaller value than that of the ZFC M-H loops at low temperature and it increased with the increasing temperature. The $H_C$ of the +50 kOe FC M-H loops showed a maximum value at ~25 K before merging with the ZFC $H_C$ value as the temperature increased.

Fig. 7 (c) and (d) show the cooling field ($H_{FC}$) dependence of the exchange bias shift ($H_E$) and the coercivity ($H_C$) of the film, measured at 5 K. The exchange bias shift showed a large increase for $H_{FC}$ = 5 kOe as compared to the ZFC value. However as $H_{FC}$ increased beyond 5 kOe, the exchange bias field ($H_E$) decreased almost monotonically and a lower than ZFC EB shift was observed for $H_{FC} = 50$ kOe. While the coercivity ($H_C$) of the film decreased rapidly with the increasing cooling field for $H_{FC} \leq 10$ kOe and as the $H_{FC}$ increased beyond 10 kOe it shows almost a constant value. Previously, Wang et al.[7] and Nayak et al.[8] had also showed that the coercivity of the bulk ZFC EB systems decreased in the FC measurements. However, it is known that in CEB systems the coercivity generally increased in the FC measurements.[28]

### c. Zero Field Cooled (ZFC) training effect

Another important feature of the exchange bias effect is training effect. The training effect is extensively used to understand the exchange coupling behaviour at the interface of the conventional exchange bias systems.[31, 33-35] I have also studied the training effect in this Mg-ferrite thin film to understand the origin of the ZFC EB effect in the system. Here unlike the CEB systems (in CEB systems, training effect is studied in Field cooled mode), the film was cooled down to 10 K from RT without a magnetic field. Then consecutive training M-H loops were collected by sweeping the magnetic field at 10 K. Fig. 8 (a) shows low field part these M-H loops. The complete M-H loops are shown in the inset of Fig. 8 (a). The Fig. 8 (a) clearly shows that the exchange bias shift ($H_E$) and coercivity ($H_C$) of the M-H loops decrease as a result of consecutive M-H loop iterations (loop number 'n'). Similar behaviour is also observed in Conventional Exchange Bias (CEB) systems.[30, 36] Though the CEB systems necessarily require to field cool before the training measurements.[30, 36] It is also interesting to note that the remanence magnetizations (both $|M_{r1}|$ and $|M_{r2}|$) of the film increased with the increasing 'n'. Whereas in case of the CEB systems, the training effect of the M-H loop (that shifted along negative field axis), generally shows a decrease in the $|M_{r1}|$ value with increasing 'n'.[32, 34, 37, 38] While the $|M_{r2}|$ has both decreasing and increasing tendencies depending on the CEB systems.[34, 37-39]

The decrease in the exchange bias field ($H_E$) of the training M-H loops were extensively studied in different CEB systems[30, 33, 36] and most of them follow an empirical power law relation

$$H_E - H_{E\infty} = \frac{K_E}{\sqrt{n}} \qquad (1)$$

Where, $H_E$ is the exchange bias field for the nth M-H loop, $H_{E\infty}$ is the EB field for $n = \infty$ and $K_E$ is a proportionality constant. This behaviour was attributed to the thermodynamic relaxation of the interfacial spins and it is found that most of the CEB systems obey this behaviour for n > 1. [32, 40] The $H_C$ and the $M_E$ ($=\frac{|M_{r2}|-|M_{r1}|}{2}$) of the training M-H loops of these CEB systems also show similar trend for n >1. [30, 40, 41]

Mishra et al.[35] had proposed another mechanism for the training effect. They had considered the frozen spin relaxation and the spin rotation at the interface of the CEB systems during training effect measurements and the exchange bias shift was formulated as [35]

$$H_E = H_{E\infty} + A_f e^{-\frac{n}{P_f}} + A_i e^{-\frac{n}{P_i}} \qquad (2)$$

Where, $A_f$ and $P_f$ are the parameters related to the frozen spin relaxation, $A_i$ and $P_i$ are the parameters related to the spin rotation. The A factors are the weight factor and have the dimension of magnetic field, the P is a dimensionless parameter related to relaxation rate. [35, 40, 42, 43]

The exchange bias field ($H_C$) and coercivity ($H_C$) of the Mg-ferrite thin film are plotted as a function of the training loop index number 'n' in the Fig. 8 (b). Fig. 8 (c) and (d) show the $M_E$ and average $M_r$ of the film with 'n', respectively. The exchange bias field ($H_E$) can be fitted with the equation 1 for n > 1. Whereas the $H_E$ of the film shows good fitting with only one exponent of equation 2 for all 'n'. Similar to the $H_E$, the $H_C$, $M_E$ and $M_r$ of the film are also fitted with the equation 1 for n > 1 and with one exponent of equation 2 for all 'n'. The dimension and notation of the parameters of the equations 1 and 2 are changed accordingly for the fitting of $H_C$, $M_E$ and $M_r$. Table 1 shows the parameters obtained from the fittings of exchange bias field, $H_E$. The fitting with equation 1 for $H_E$ yield $H_{E\infty}$ = -50 Oe. Here one needs to note that, previously the negative $H_{E\infty}$ was also obtained in Fe$_3$O$_4$ film.[40] However, the sign change in EB shift was not observed even after large number of M-H loop iterations. Therefore, the sign change in this Mg-ferrite thin film is also not anticipated and rather it is likely that the M-H loop might become symmetric after a large number of loop iterations since the loop shift become very small ($H_E$ ~ 20 Oe) for n = 6. This indicates that the thermal relaxation of the interfacial spins might not be a feasible explanation of the ZFC training effect of Mg-ferrite thin film. On the other hand, the fitting with one exponent of equation 2 provide the 'P' factor (P = 0.845) value similar to the relaxation rate of the frozen spins ($P_f$) obtained in different CEB systems.[35, 43] This indicates that the ZFC training effect of the film is most likely to be dominated by the relaxation of the frozen spins at the interface of the grain boundary. Nevertheless, a good fitting with one exponent of the equation 2 for the training $H_C$, $M_E$ and $M_r$ of the film also support the frozen spin relaxation behaviour.

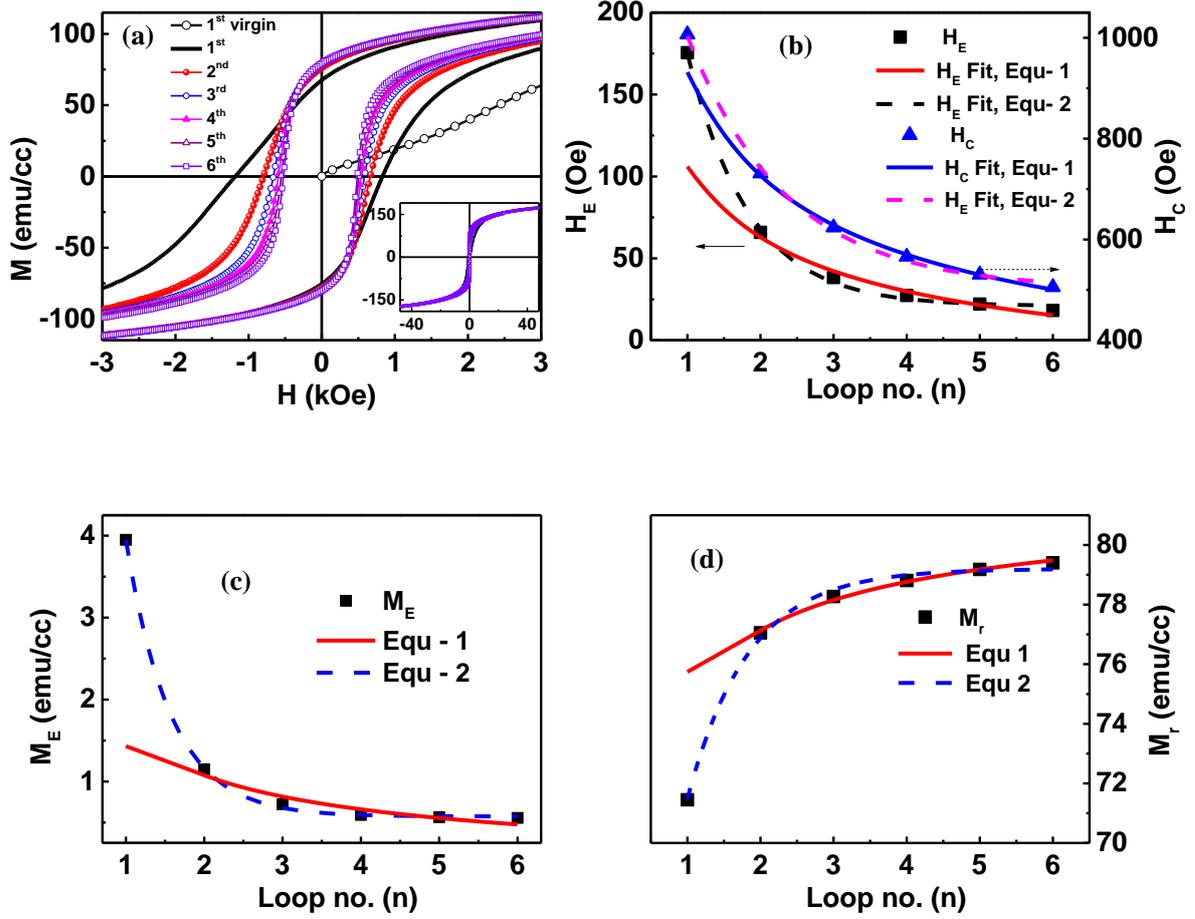

Fig. 8. (a) Zoomed view of the ZFC training M-H loops at 10 K. Inset: Full M-H loops (for n = 1 and 6). (b) Exchange bias Field ($H_E$) and coercivity ($H_C$) as a function of training M-H loop iteration number n. Both $H_E$ and $H_C$ are fitted with equation 1 and 2. (c) $M_E$ and (d) $M_r$ with n.

Table 1. Parameters obtained from the fitting of training effect

| From equation (1) | | From equation (2) | | |
|---|---|---|---|---|
| $K_E$ (Oe) | $H_{E\infty}$ (Oe) | $H_{E\infty}$ (Oe) | $A_f$ (Oe) | $P_f$ |
| 159 | −50 | 20 | 503 | 0.845 |

## IV. Discussion

The conventional exchange bias effect is generally attributed to the exchange coupling between the interfacial spins of two magnetic materials such as ferromagnetic (FM) – Antiferromagnetic (AFM)[1, 36], FM- spin glass[44, 45], FM – Ferrimagnetic (FIM)[46, 47] etc. There are also some single phase (crystallographic phase) materials that show CEB effect.[30, 41, 48] However, these single phase materials show coexistence of different magnetic orders within them and the exchange coupling at the interface of these magnetic orders resulted in exchange bias effect.[30, 41, 48] The XRD of our thin film shows single phase of Mg-ferrite cubic spinel structure. The XPS data also supported it, since no impurity element was found. Therefore to understand the exchange bias phenomenon in this film one needs to know the magnetic orderings within it. The thermomagnetic measurements were performed to address this. Fig. 9 (a) and (b) show the thermomagnetic data (M-T) of the film. The M-T were measured in both Zero Field Cooled (ZFC) and Field Cooled (FC) modes (in Fig. 9 (a)). We can see that the ZFC M-T data deviates from the FC M-T data at $T_{irr}$ (indicated with arrow in the Fig. 9 (a)).

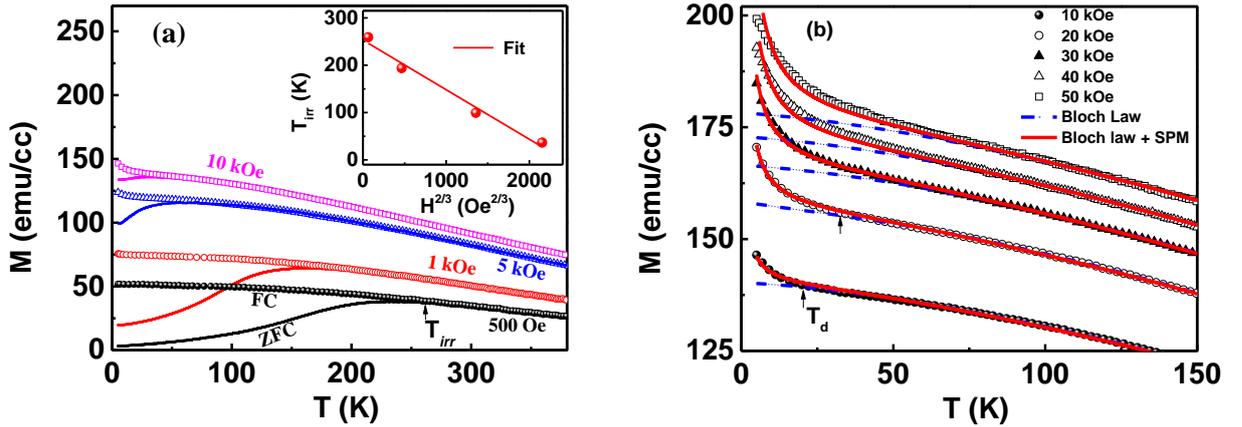

Fig. 9. (a) ZFC and FC M-T data of the Mg-ferrite thin film. Inset: $T_{irr}$ vs $H^{2/3}$, the line represents Thouless and de Almeida fitting. (b) High field FC M-T data. The blue dashed lines are the fitted data using Bloch's law for ferrimagnetic sample and red lines are due to coexistence of ferrimagnetic and SPM grains in the film.

This behaviour is generally attributed to the freezing of the moments of smaller grains.[45, 49] Below $T_{irr}$ the spin of the smaller nanocrystalline grains frozen to the random direction as the crystalline anisotropy of the grains overcome the thermal fluctuation.[49] The value of $T_{irr}$ decreases as the applied magnetic field increases. The decrease of $T_{irr}$ with the increasing magnetic field followed the famous Thouless and de Almeida line[50, 51] ($T_{irr} \propto H^{2/3}$) for spin glass systems, shown in the inset of the Fig. 9 (a).

The high field FC M-T data is shown in the Fig. 9 (b). These FC M-T data shows good fit with the Bloch's law[52] ($M(T) = M_0 \left(1 - BT^{\frac{3}{2}}\right)$), for temperature dependence magnetization of a ferrimagnetic system, above $T_d$ (indicated by arrow). As temperature decreased below $T_d$, an upturn in the magnetization is observed. It is also observed that the value of $T_d$ increased with the applied magnetic field. I assumed that this behaviour is due to the coexistence of ferrimagnetic and superparamagnetic (SPM) grains in the film (formulated as $M(T) = M_0^* \left(1 - BT^{\frac{3}{2}}\right) + C^*/T$, where $M_0^* = (1-x)M_0$ and $\frac{C^*}{T} = \frac{xCH}{T}$, where 'x' is the volume fraction of the superparamagnetic state, $C$ is the Currie constant of the superparamagnetic state). The 10 – 30 kOe FC M-T data show good fit with this assumption. However, as the field increased (above 30 kOe), the low temperature data shows a tendency towards saturation as compared to the fitted data. This behaviour could be due to a weak ordering of the SPM grains under application of high magnetic field. The similar field induced ordering of the SPM grains were also predicted in different ZFC EB systems.[7, 25] Therefore, it is likely that the smaller grains of this Mg-ferrite nanocrystalline thin film were frozen into a spin glass like state as the temperature decreased much below the $T_{irr}$. The pinning effect of the surface spins of these frozen glassy states at the interface of ferrimagnetic grains could possibly leads to the observed exchange bias effect. Earlier, exchange bias effect was also reported in different single phase ferrite thin films such as Ni-ferrite, MnZn-ferrite thin films.[16-18] The observed EB effect in these systems was also speculated due to the pinning effect of the surface spins of a disordered state (glassy states/super paramagnetic (SPM) states) at the interface of the ferrimagnetic grains. However, softening (decrease of $H_C$) of the FC M-H loops could be associated with the field induced weak ordering of the SPM grains. The ordering within these SPM grains reduced the net anisotropy of the system. I speculate similar effect in the training effect measurements, here the frozen SPM grains relaxed towards an ordered state and decreased the anisotropy of the system. This reduced anisotropy leads to a square trained M-H loop (or increase in $|M_{r1}|$ and $|M_{r2}|$) as compared to the initial ZFC M-H loop.

# V. Conclusion

A single layer Mg-ferrite thin film was deposited on amorphous quartz substrate using pulsed laser ablation technique. This film showed ZFC EB effect along with the ZFC training effect. The film also showed CEB effect in field cooled measurements. The observed exchange bias effect is attributed to the pinning effect of the surface spins of frozen glassy states at the interface of ferrimagnetic grains. The decrease in the coercivity of the field cooled M-H loop is speculated due to a weak field induce ordering of the superparamagnetic grains.

## Acknowledgements

I thank Prof. Shiva Prasad and Prof. N. Venkataramani for the lab facilities. I also thank SAIF and IRCC of IIT Bombay for the VSM, XRD, ESCA and FEG-SEM facilities.